\documentclass[conference]{IEEEtran}
\IEEEoverridecommandlockouts

\newcommand{\mli}[1]{\mathit{#1}}
\usepackage[inkscapeformat=png]{svg}
\usepackage{bbm}
\usepackage{multirow}
\usepackage{hyperref}
\hypersetup{nolinks=true}
\usepackage{cite}
\usepackage{amsmath,amssymb,amsfonts}
\usepackage{algorithmic}
\usepackage{graphicx}
\usepackage{textcomp}
\usepackage{xcolor}
\usepackage{tikz}
\def\BibTeX{{\rm B\kern-.05em{\sc i\kern-.025em b}\kern-.08em
    T\kern-.1667em\lower.7ex\hbox{E}\kern-.125emX}}

\newcommand\copyrighttext{%
  \footnotesize \textcopyright This work has been submitted to the IEEE for possible publication. Copyright may be transferred without notice, after which this version may no longer be accessible.}
\newcommand\copyrightnotice{%
\begin{tikzpicture}[remember picture,overlay]
\node[anchor=south,yshift=10pt] at (current page.south) {\fbox{\parbox{\dimexpr\textwidth-\fboxsep-\fboxrule\relax}{\copyrighttext}}};
\end{tikzpicture}%
}  

\begin{document}
\bstctlcite{IEEEexample:BSTcontrol}
\title{Rural Handover Parameter Tuning to Achieve End to End Latency Requirements of Future Railway Mobile Communication Systems \\
}

\author{\IEEEauthorblockN{Dogukan Atik, Murat Gursu}
\IEEEauthorblockA{\textit{Nokia Standards} \\
\textit{Nokia Solutions and Networks GmbH}\\
Munich, Germany \\
name.surname@nokia.com}
\and
\IEEEauthorblockN{Fidan Mehmeti, Wolfgang Kellerer}
\IEEEauthorblockA{\textit{Chair of Communication Networks} \\
\textit{Technical University of Munich}\\
Munich, Germany \\
name.surname@tum.de}
}

\maketitle
% *** IEEE Copyright notice with TikZ ***
% 
\copyrightnotice

\begin{abstract}
GSM-R (GSM for Railways) is a 2G-based standardized ground-to-train communications system that enabled interoperability across different countries. However, as a 2G-based system, it is nearing its lifetime and therefore, it will be replaced with 5G-based Future Railway Mobile Communications System (FRMCS). FRMCS is expected to bring in new use cases that demand low latency and high reliability. However, from a mobility perspective, it is not clear how the low latency and high reliability will be achieved. This paper investigates the effect of handover procedure on latency and reliability and analyzes which use cases of FRMCS can be satisfied using baseline handover. We also sweep through different handover parameter configurations and analyze their effect on mobility performance. Then, we analyze the effect of mobility performance on packet latency and reliability. Our results show that, with baseline handover, Standard Data Communications Scenario is met and optimizing for baseline handover performance can reduce latency by up to $18.5$\%, indicating that optimizing for mobility performance is crucial in FRMCS.
\end{abstract} 

\section{Introduction}
\label{sect:intro}
GSM for Railways (GSM-R) is a 2G-based, standardized ground-to-train communication system that enables interoperability across different country borders. In addition to interoperability, it offers other services, such as voice, data services and perhaps most notably, railway emergency call~\cite{eirene_frs}. Over time, it is deployed significantly throughout Europe and in the world, and today nearly $150,000$\:km of railway lines in Europe (and over $250,000$\:km in the rest of the world) operate using GSM-R~\cite{frmcs_5g_rail_uic_brochure}.

However, GSM-R is expected to be obsolete by 2035~\cite{uic_frmcs}, and this resulted in a renovation effort led by the International Union of Railways (UIC). The successor of GSM-R bears the name Future Railway Mobile Communications System (FRMCS). It is a 5G-based railway communication system with its own dedicated frequency bands, a paired band at $900$\:MHz and an unpaired band at $1900$\:MHz.

FRMCS will serve a pivotal role in railway digitalization by enabling new applications, such as train automation, self-driving trains, remote control and monitoring of on-board train equipment~\cite{frmcs_5g_rail_uic_brochure}. Of course, these foreseen applications come up with stringent reliability and latency requirements~\cite{3gpp.22.289}. Table~\ref{tab:scenarios_and_reliability} lists the use cases together with their respective latency, reliability and throughput requirements.

\begin{table*}[!htb]
\caption{List of Railway Communication Scenarios and Their Requirements~\cite{3gpp.22.289}}
\setlength\tabcolsep{10pt}
\begin{center}
\begin{tabular*}{\textwidth}{|c|c|c|c|c|} %{|c|c|c|c|c|}
\hline
\textbf{Scenario}&{\textbf{End-to-end}}&{\textbf{Reliability}}&{\textbf{Speed Limit}}&{\textbf{User Experienced}} \\
                 &{\textbf{latency}}   &                      &                      &{\textbf{Data Rate}} \\
\hline
Voice Communication for operational purposes & $\leq$ 100 ms & 99.9\% & $\leq$ 500 km/h & 100 kbps up to\\
                                             &               &        &                 & 300 kbps \\
\hline
Critical Video Communication for observation purposes & $\leq$ 100 ms & 99.9\% & $\leq$ 500 km/h & 10 Mbps \\
\hline
Very Critical Video Communication with a direct impact on train safety & $\leq$ 100 ms & 99.9\% & $\leq$ 500 km/h & 10 Mbps up to \\
                                                                       &               &        &                 & 20 Mbps\\
\cline{2-5}
                             & $\leq$ 10 ms  & 99.9\% & $\leq$ 40 km/h  & 10 Mbps up to\\
                             &               &        &                 & 30 Mbps\\
\hline                             
Standard Data Communication  & $\leq$ 500 ms & 99.9\% & $\leq$ 500 km/h & 1 Mbps up to\\
                             &               &        &                 & 10 Mbps\\
\hline
Critical Data Communication  & $\leq$ 500 ms & 99.9999\% & $\leq$ 500 km/h & 10 kbps up to\\
                             &               &           &                 & 500 kbps\\
\hline
Very Critical Data Communication           & $\leq$ 100 ms & 99.9999\% & $\leq$ 500km/h  & 100 kbps up to\\
                                           &               &           &                 & 1 Mbps\\
\cline{2-5}
                             & $\leq$ 10 ms  & 99.9999\% & $\leq$ 40 km/h  & 100 kbps up to\\
                             &               &           &                 & 1 Mbps\\
\hline
Messaging                    &      -        & 99.9\%    & $\leq$ 500 km/h & 100kbps\\
\hline
\end{tabular*}
\label{tab:scenarios_and_reliability}
\end{center}
\end{table*}
  
Reliability in FRMCS is defined as the percentage of network layer packets that are successfully transmitted within the latency constraint (end-to-end latency). For example, only one packet lost in one million sent packets can be tolerated for the very critical data communication scenario, where the train speed can go up to $500$\:km/h~\cite{3gpp.22.289}.

Handovers can have an adverse effect on reliability since they can increase the latency significantly. Considering the handover interruption time is in the order of $40$\:ms - $60$\:ms for L3 handover~\cite{5gsmart_deliverable_d1_5}, it is a significant source of latency, especially in the scenarios where the end-to-end latency is bounded by $100$\:ms. During the handover interruption time, since the UE is not connected to any cell, there is no transfer of uplink/downlink (UL/DL) user data. Therefore, the packets wait in the queue and experience the latency of interruption time. For this reason, the effect of mobility on packet latency, and subsequently reliability, should be clarified.

Our contributions can be summarized as follows:
\begin{itemize}
\item We investigate the effect of handover interruption time on reliability by conducting simulation studies that are in line with 3GPP reference FRMCS scenarios~\cite{etsi.103.554-2}.
\item We sweep through different handover parameter values and discuss how they affect mobility performance and subsequently, packet latency and reliability.
\item We analyze which of the FRMCS scenarios in Table~\ref{tab:scenarios_and_reliability} can be met with L3 handover procedure.
\end{itemize}

The remainder of the paper is organized as follows: Section~\ref{sect:related_work} discusses the related work in the literature. Section~\ref{sect:system_model} explains the sources of interruption the UE experiences. Section~\ref{sect:simulation_scenario} introduces the simulation scenario that is inline with the 3GPP standards. Section~\ref{Sect:Results} elaborates on the results obtained with L3 handover parameter tuning and which scenarios are met. Section~\ref{sect:future_outlook} discusses some possible future research directions and Section~\ref{sect:conclusion} concludes the paper.

\section{Related Work}
\label{sect:related_work}
To best of our knowledge, there is no prior work that investigates the effect of handover interruption time on packet latency and reliability in FRMCS. Reference~\cite{noh2020high} discusses how to reduce the interruption time using different mobility procedures in a high speed rail scenario. In~\cite{ali2020seamless}, a frequency switch scheme is proposed that reduces the interruption time when the train moves between different remote antenna units that belong to the same central unit (CU). References~\cite{wu2022parameter},~\cite{chen2022handover} are examples of handover parameter optimization algorithms for high-speed rail. Although these works describe some methods to reduce the interruption time, they don't evaluate the effect of reducing interruption time on reliability, which is the focus of our work.

The effect of handover on TCP and UDP traffic is shown in~\cite{karamichailidis2021session} and~\cite{cvetkovski2022railway}. However, the handover test setup presented in these papers do not include UE mobility. Additionally, the UE speed at the field test in~\cite{cvetkovski2022railway} is $12$\:km/h, too low when considering the train speeds up to $500$\:km/h. 
Reference~\cite{5grail_deliverable_d6_2} delivers the results of a network simulator developed for FRMCS, but the handover mechanism is not 5G-compliant and the effect of handovers is reported for Wi-Fi AP's, therefore is not representative for our purposes. 

\section{System Model}
\label{sect:system_model}
End to end latency of an application includes the latency of the packets over the wired link (processing, transmission, propagation and queuing delays)~\cite{bovy2002analysis} and the wireless link. In our studies, the wireless link between the train and the FRMCS application is the link between the train and the gNodeB. From a broader perspective, our motivation is to address the mobility challenges listed in~\cite{atik2023reliabilityfrmcs}, and specific to this study, our motivation is to analyze the effect of handover on latency and optimize for it. For these reasons, our focus is on the latency of the network layer packets over the radio and the rest of the discussion is carried out with this in mind. 

\begin{figure}[t!]
\centerline{\includegraphics[width=0.4\textwidth]{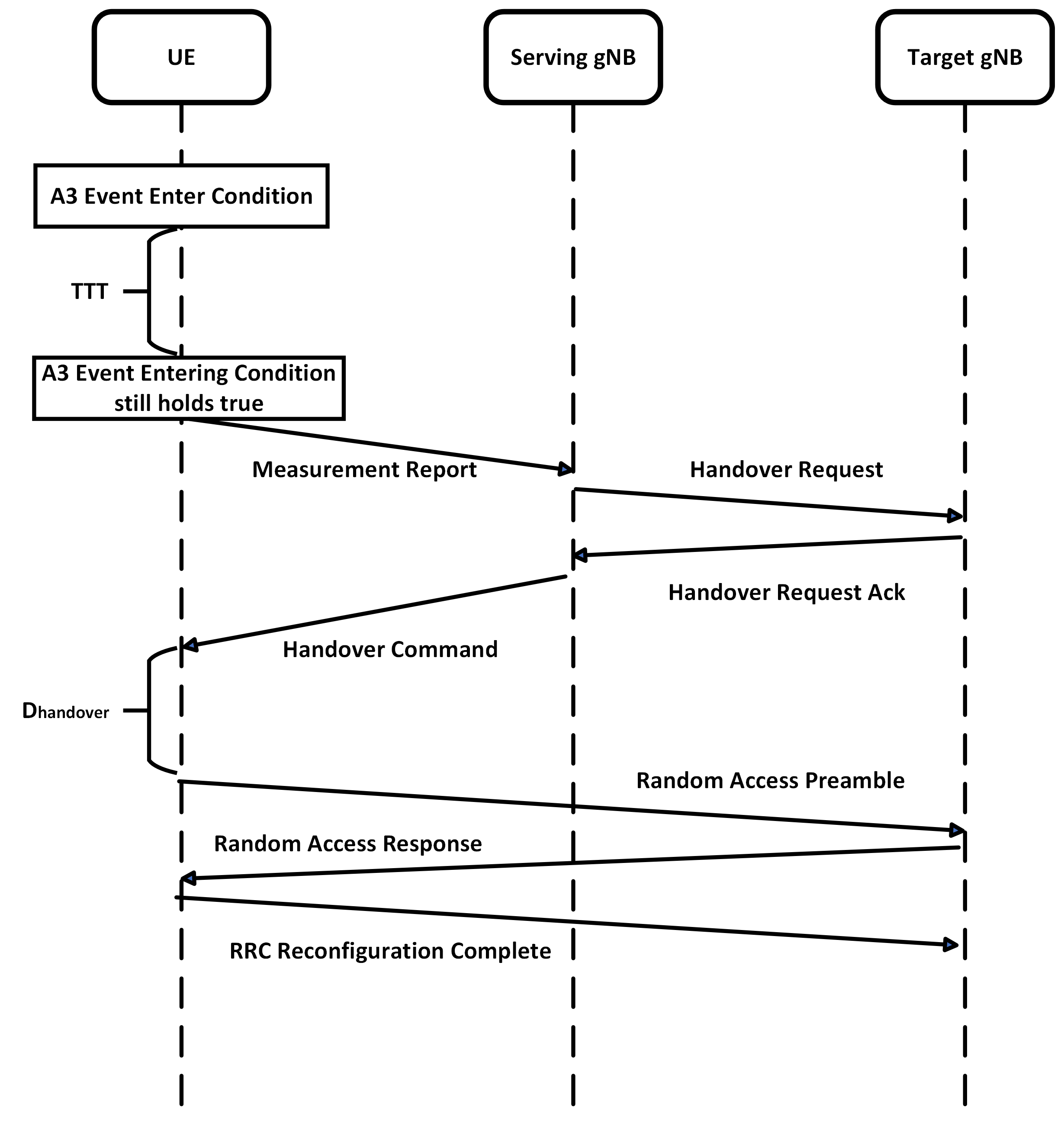}}
\caption{Baseline Handover Procedure.}
\label{fig:BHO}
\end{figure}

The UE periodically measures reference signals in the specified frequency band and compares the L3 filtered RSRP measurements of the serving cell and any other identified neighboring cells to check whether A3 Event Entering condition\footnote{The measurement events in 3GPP TS 38.331 are defined with leaving and entering conditions to make it state like. Once an entering condition is triggered UE is in event entered state.} \cite{3gpp.38.331} is satisfied:
\begin{equation}
M_n + O_{fn} + O_{cn} - H > M_{p} + O_{fp} + O_{cp} + O, \label{eq:A3EventEnter}
\end{equation}
where $M_n$ ($M_p$) is the measurement of the neighboring (serving) cell, $O_{fn}$ ($O_{fp}$) and $O_{cn}$ ($O_{cp}$) are the frequency-band-specific and cell-specific offsets of the neighboring (serving) cell, $O$ is the offset parameter defined for the event, and $H$ is the hysteresis\footnote{Hysteresis parameter behaves in a directional manner to avoid UE ping ponging between event entering and event leaving states.} parameter for this event. In our studies, all cell-specific and frequency-band-specific offset parameters are set to zero. Therefore after re-arranging, \eqref{eq:A3EventEnter} simplifies to
\begin{equation}
M_n - M_p > O + H. \label{eq:A3EventEnterSimplified}
\end{equation}
Once the A3 Event Entering condition is satisfied, the UE starts a \textit{Time-to-Trigger} (TTT), and during the TTT, it checks whether the A3 Event Leaving Condition ($ M_n - M_p < O - H$) is satisfied. If during this time period, the leaving condition is not satisfied, then the event is considered triggered and the UE sends the measurement report to the serving cell. After receiving the measurement report, the serving gNodeB sends handover request to the target gNodeB (in case the target cell is controlled by another gNodeB) and acquires target cell configuration so that the UE can perform a handover to the target cell. Then, the serving gNodeB indicates the UE that it can perform handover via the Handover Command. After this, the UE disconnects from the serving cell and initiates the random access procedure to the target cell to establish UL synchronization with the target cell. After the random access is successful, the UE indicates the success of the handover to the new serving cell with the RRC Reconfiguration Complete message. Fig.~\ref{fig:BHO} visualizes the flow of incidents described.

The sources of latencies in Fig.~\ref{fig:BHO} are TTT, the time spent during the exchange of Handover Request and Handover Request Ack messages, the time spent in random access, and $D_{handover}$~\cite{3gpp.38.133}, which is the delay between the reception of the Handover Command and the transmission of the first random access preamble. TTT is a handover parameter that is studied in Section~\ref{Sect:Results} and the time spent during the exchange of Handover Request and Handover Request Ack messages is modelled as a delay since its details are beyond the scope of this paper. 

\begin{figure}[t!]
\centerline{\includegraphics[width=0.4\textwidth]{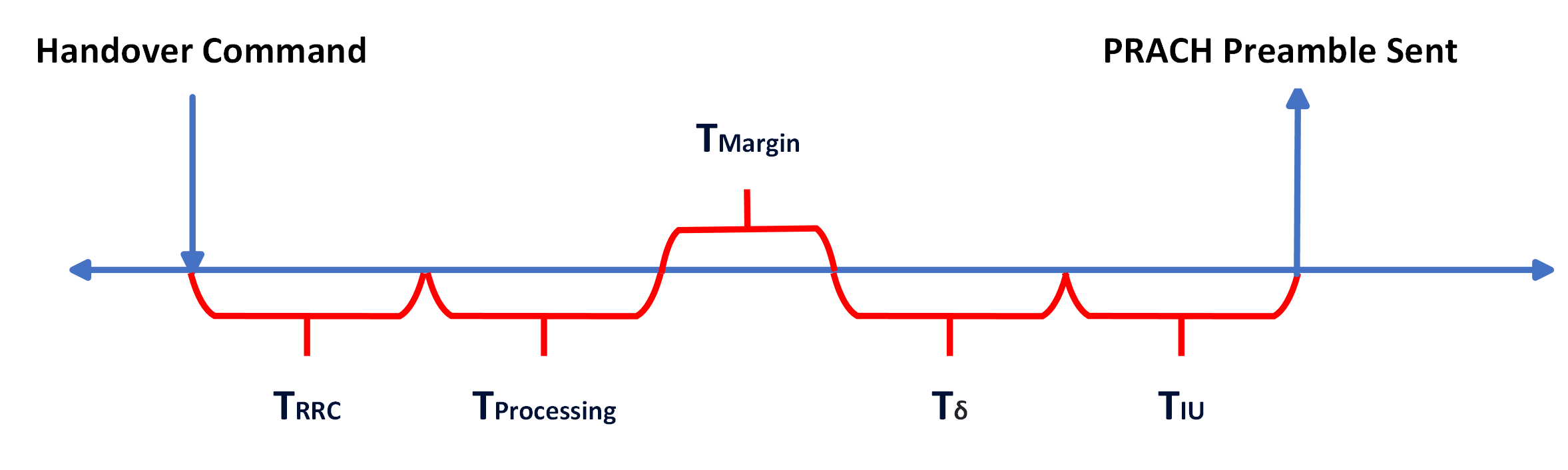}}
\caption{Time spent between the reception of Handover Command and the transmission of the first PRACH Preamble.}
\label{fig:DHandover}
\end{figure}

In Fig.~\ref{fig:DHandover}, the components of $D_{handover}$ are illustrated. $T_{RRC}$ corresponds to the RRC procedure delay related to receiving and decoding the Handover Command and the target cell configuration as described in~\cite{3gpp.38.331}. During this time, the UE is still connected to the serving cell. Then, the UE disconnects from the source cell in an attempt to connect to the target cell. $T_{processing}$ is defined as the time for UE processing as in applying the target cell configuration~\cite{3gpp.38.133} and this is set to $20$\:ms as indicated in~\cite{3gpp.38.331}. Similarly, $T_{margin}$ is the time for SSB post-processing and it is set to $2$\:ms. $T_{\delta}$ is the time required for fine time tracking and its value is taken as $5$\:ms, since it is assumed that the UE is not provided with an SMTC configuration. $T_{IU}$ is the uncertain amount of time the UE spends until acquiring the first available Physical Random Access Channel (PRACH) occasion to send the preamble to the target cell. It is smaller than or equal to the summation of SSB to PRACH occasion association period~\cite{3gpp.38.213} and $10$\:ms. Since the association period is $10$\:ms in our setup, $T_{IU}$ can be up to $20$\:ms. Further detail on these delay sources can be found in~\cite{3gpp.38.133}. 

\section{Simulation Scenario}
\label{sect:simulation_scenario}
\begin{figure}[t!]
\centerline{\includegraphics[width=0.4\textwidth]{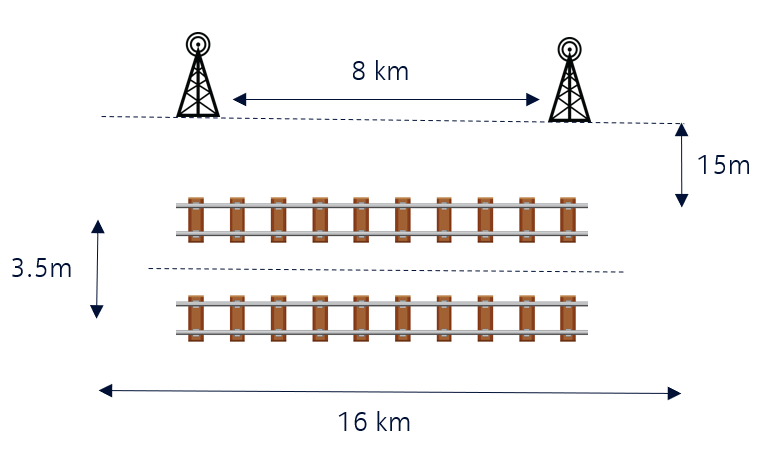}}
\caption{Schematic of the railway track used in the simulation studies.}
\label{fig:railway_track}
\end{figure}

The simulation studies are carried out using a system-level simulator developed inside Nokia~\cite{freac}. Reference~\cite{etsi.103.554-2} foresees three different scenarios for railway operation: urban, rural and hilly scenarios. In our simulations, the focus was on the rural scenario since it allowed for higher train speeds compared to the other settings. Sticking with the rural scenario setup, the inter-site distance (ISD) is set to $8$\:km. Fig.~\ref{fig:railway_track} shows the layout of the railway track used in the simulation studies. All the parameters are from~\cite{etsi.103.554-2} except the railway track length, which is $16$\:km. On these tracks, two trains moving in opposite directions with the speed of $500$\:km/h are simulated. The starting point of the trains on the railway track is randomized in each simulation run. The simulation duration is $115.14$ seconds, which is the time it takes for the trains to travel $16$\:km. The gNodeB and train antenna height, which are not shown in Fig.~\ref{fig:railway_track}, are listed in Table~\ref{tab:simulation_parameters}.

\begin{table}[htbp]
\caption{Simulation Parameters \cite{etsi.103.554-2}}
\begin{center}
\begin{tabular}{|c|c|}
\hline
\textbf{Parameter}&{\textbf{Value}} \\
\hline
Carrier Frequency & 900 MHz \\
\hline
Carrier Bandwidth & 3 MHz \\
\hline
Subcarrier Spacing & 15 kHz \\
\hline
Transmission Mode & FDD \\
\hline
Channel Model & TR38.901 Rural Macro \\
 & with Stochastic LOS \\
\hline
gNodeB Height & 35 meters \\
\hline
Train Antenna Height & 4 meters \\
\hline
Traffic Model & FTP Model 3 \\
\hline
FTP Packet Size & 0.5 MB \\
\hline
FTP Packet Arrival Rate & 2.5 \\
(packets per second) & \\
\hline
Simulation realizations & 29 \\
\hline
\end{tabular}
\label{tab:simulation_parameters}
\end{center}
\end{table}

Table~\ref{tab:simulation_parameters} lists the rest of the important simulation parameters. The channel model is from the 3GPP standards and it is described in~\cite{3gpp.38.901}. The pathloss and LOS probabilities are calculated based on the Rural Macro scenario. Moving on to the carrier frequency, FRMCS supports operation on two different frequency bands, namely the $900$\:MHz FDD band with $5.6$\:MHz of DL/UL bandwith, or $1900$\:MHz TDD band~\cite{ecc_20_02}. A migration period where GSM-R is gradually replaced with the FRMCS equipment is foreseen approximately between 2025 and 2035~\cite{uic_frmcs}. During this migration period, the FRMCS equipment and the GSM-R equipment are expected to coexist and share the same frequency band, since GSM-R also uses the $900$\:MHz band~\cite{etsi.gsm.05.01}. Due to this coexistence, we assumed the FRMCS system uses $3$\:MHz of bandwidth, while the remaining $2.6$\:MHz is reserved for GSM-R operation.  

The reliability requirements listed in Table~\ref{tab:scenarios_and_reliability} have different traffic load characteristics for different use cases. In our studies, we kept the load at $10$\:Mbps for both uplink and downlink, since it is the highest load in data communication use cases in Table~\ref{tab:scenarios_and_reliability}. FTP Model 3 is used to model the data communication between the UE (train) and the gNodeB. It is characterized by a constant file size of $0.5$\:MB and the packet arrival process is Poisson~\cite{3gpp.36.889}. To match the $10$\:Mbps load, we set the packet inter-arrival rate to be $2.5$ packets per second. The simulations are run for 12 different A3 Event Offset ($O$), TTT pairs, where $O \in \{2, 4, 6, 8\}$ dB and TTT $\in \{80, 160, 240\}$ ms. For each scenario, the number of simulation realizations was 29. On average, we generated $11.600.000$ network layer packets per scenario. Then, the empirical packet latency distribution of each scenario is constructed using the generated packets and the $99.9$th percentile of the constructed latency distributions are evaluated in Section~\ref{Sect:Results}.

\section{Results}
\label{Sect:Results}
Throughout the simulations, the following reliability definition is used:
\begin{equation}
\label{eq:reliability}
    r(L) = \frac{1}{N}\sum_{i=1}^{N}\mathbbm{1}\{l_i \leq L\} \times 100,
\end{equation}
where $L$ is the target latency (e.g. $500$\:ms for \textit{Standard Data Communication}), $N$ is the total number of packets across all simulation realizations for a given ($O$, TTT) pair, $l_i$ is the latency of packet $i$ and $\mathbbm{1}\{.\}$ is the indicator function that returns $1$ if the condition inside it is true, and $0$ otherwise. In short, $r(L)$ returns the percentage of the packets that have a latency lower than or equal to $L$.

\begin{figure}[t!]
\centerline{\includegraphics[width=0.4\textwidth]{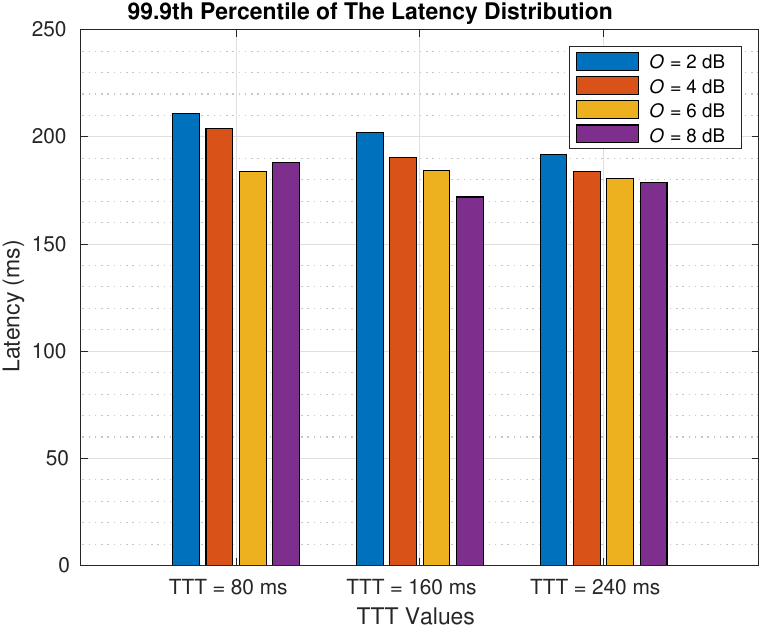}}
\caption{$99.9$th percentile of the latency distribution for different A3 Event Offset (\textit{O}) and TTT configurations.}
\label{fig:reliability}
\end{figure}

To understand the effect of handover parameters on reliability and to see for which scenarios in Table~\ref{tab:scenarios_and_reliability} the reliability and latency requirements are  met, we swept over a number of different A3 event offsets ($O$ in Eq.~\eqref{eq:A3EventEnterSimplified}) and TTT values. Then, we evaluated $99.9$th percentile of the latency distribution. The results are shown in Fig.~\ref{fig:reliability}. 

\begin{figure}[t!]
\centerline{\includegraphics[width=0.4\textwidth]{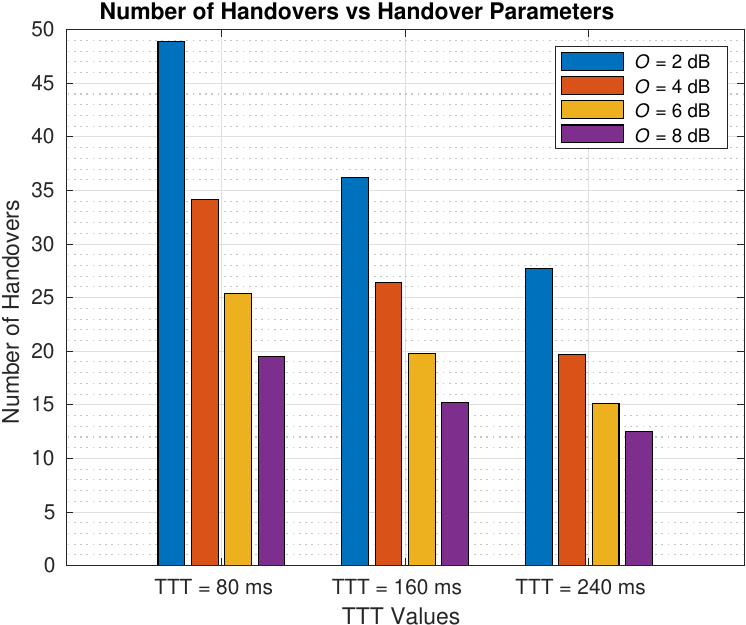}}
\caption{Total number of handovers vs. different A3 Event Offset (\textit{O}) and TTT configurations.}
\label{fig:allhandovers}
\end{figure}

Increasing A3 Event Offset reduces the number of handovers since it is harder to do a handover with a higher offset. This relationship can be observed in Fig.~\ref{fig:allhandovers}. For a fixed TTT value, increasing the offset decreases the total number of handovers occurring in the simulations. Increasing TTT for a fixed event offset has also the same effect. By setting a larger TTT, the UE waits longer before sending the measurement report to the gNodeB. As a result, the probability of executing a handover for a momentarily bad channel measurement decreases. 

\begin{figure}[t!]
\centerline{\includegraphics[width=0.4\textwidth]{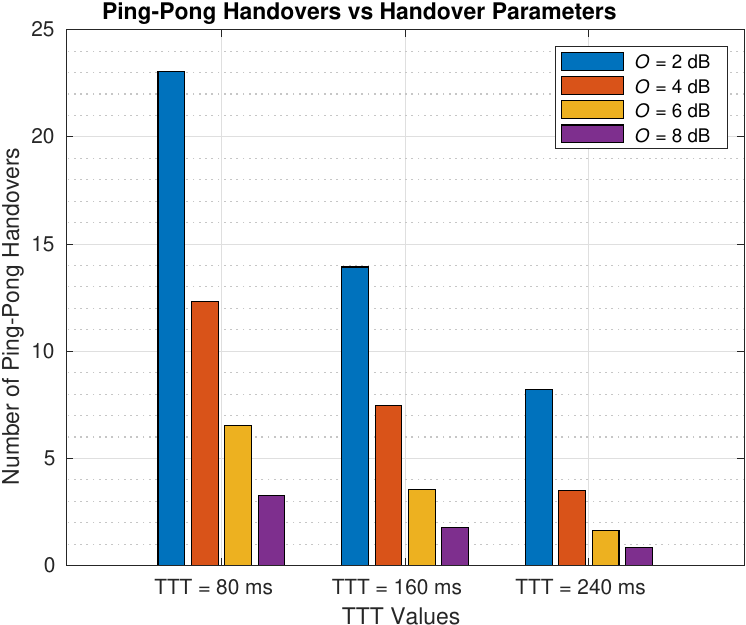}}
\caption{Number of ping-pong handovers with respect to different configurations.}
\label{fig:pingpongs}
\end{figure}

In Fig.~\ref{fig:pingpongs}, the change in ping-pong handovers with respect to the swept parameters can be observed. Increasing the A3 Event Offset decreases the likelihood of ping-pong handovers since the UE waits for the difference between the source cell and target cell measurements to be larger and it is harder to bounce back to the serving cell with a ping-pong when there is a large difference between the source and serving cell measurements. Reducing the number of ping-pong handovers is important since a significant amount of handovers in Fig.~\ref{fig:allhandovers} are due to the ping-pongs. 

A ping-pong handover in our setup is defined as a handover that is carried out to the old serving cell when the last handover from that old cell took place in less than a second ago. By reducing the number of ping-pong handovers, we can reduce the total amount of outages caused by handover interruption time, thereby reducing the latency experienced by the network layer packets due to the frequent handovers. Hence, for a fixed TTT, increasing the A3 event threshold helps with reducing the $99.9$th percentile of the latency distribution in Fig.~\ref{fig:reliability}, except for one case where TTT is $80$\:ms and \textit{O} is $8$\:dB.

\begin{figure}[t!]
\centerline{\includegraphics[width=0.4\textwidth]{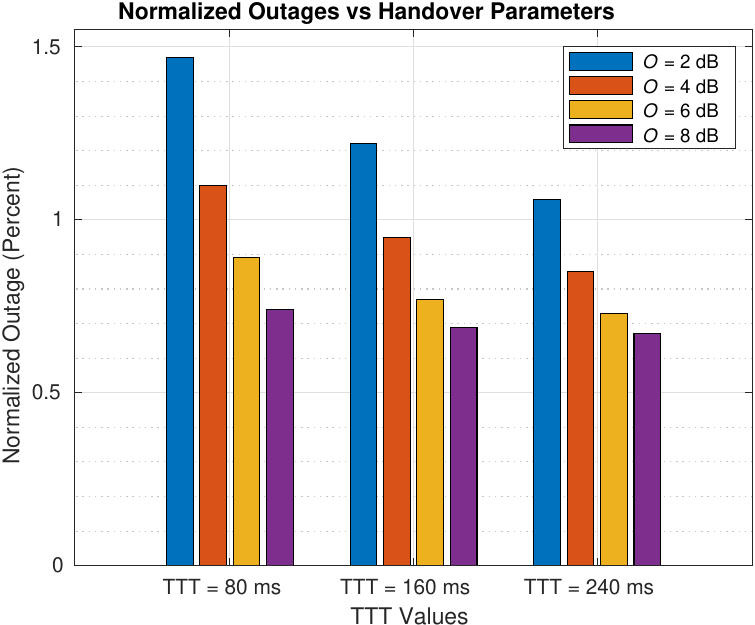}}
\caption{Normalized outage vs. handover parameters.}
\label{fig:normalizedoutage}
\end{figure}

The effect of reducing ping-pongs, or handovers in general, can be observed in Fig.~\ref{fig:normalizedoutage}, where the normalized outages with respect to the different handover parameter configurations are plotted. The normalized outage is calculated as,
\begin{equation}
    \mli{Normalized} \mli{Outage} = \frac{\sum_{k=1}^{N}o_i}{T \times K \times R} \times 100,
\end{equation}
where $o_i$ represents the duration of outage $i$, $N$ is the total number of outage instances, $T$ is the simulation duration, $K$ is the number of UEs and $R$ is the number of different simulation realizations. It can be observed that for a fixed TTT, increasing the A3 event offset decreases the normalized outage. This is because the number of handovers (and hence, the number of times a UE experiences handover outage) decreases as the A3 event offset increases. This reduction in normalized outage is also reflected in Fig.~\ref{fig:reliability}. 

\begin{figure}[t!]
\centerline{\includegraphics[width=0.4\textwidth]{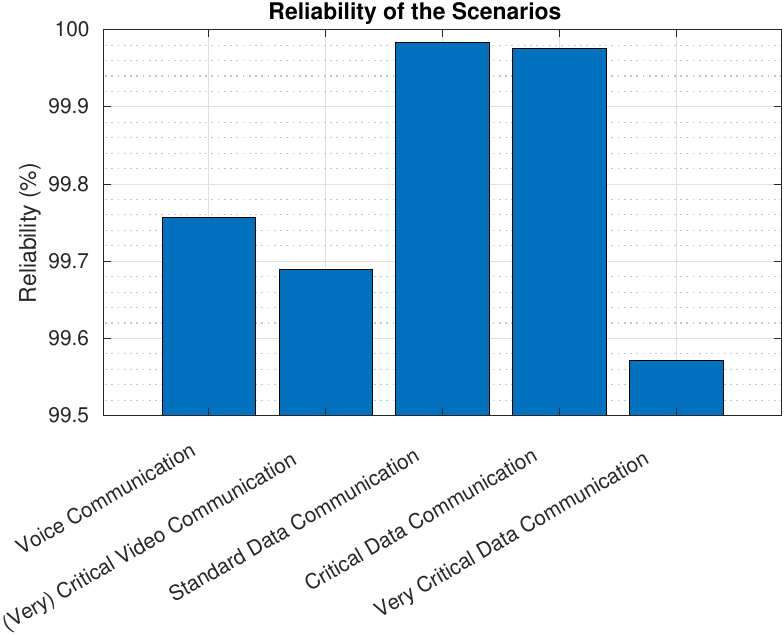}}
\caption{Reliability of the scenarios where the reliability is calculated as in~\eqref{eq:reliability} for TTT = $160$\:ms and \textit{O} = $8$\:dB}
\label{fig:reliabilityscenario}
\end{figure}

Fig.~\ref{fig:reliabilityscenario} plots the reliability of each scenario in Table~\ref{tab:scenarios_and_reliability}, except those where the train speed is lower than $40$ km/h and \textit{Messaging}. Low train speeds are considered for entering/exiting a station or maneuvering in the marshalling yards~\cite{3gpp.22.289} but we considered a fast train travelling in a rural area in our simulation studies. Therefore, cases corresponding to low speed are out of the scope of this paper. Moving to \textit{Messaging}, it has no latency requirement. Therefore, reliability (based on~\eqref{eq:reliability}) can not be calculated for it. Hence, Fig.~\ref{fig:reliabilityscenario} does not include the \textit{Messaging} scenario.

While obtaining Fig.~\ref{fig:reliabilityscenario}, the simulations were carried out using the setting that corresponds to the lowest latency at $99.9$th percentile in Fig.~\ref{fig:reliability}. Hence, TTT is set to $160$\:ms and $O$ is set to $8$\:dB. For \textit{(Very) Critical Video Communication} and \textit{Standard Data Communication} scenarios, the load is kept as $10$\:Mbps, which corresponds to the arrival rate of $2.5$ packets per second, as illustrated in Table~\ref{tab:simulation_parameters}. However, to be in line with the load requirements of the other scenarios in Table~\ref{tab:scenarios_and_reliability}, the arrival rate of the FTP3 Packets are modified such that the load is $500$\:kbps for \textit{Critical Data Communications}, $1$\:Mbps for \textit{Very Critical Data Communications} and $300$\:kbps for \textit{Voice Communications}. Finally, the load in our simulation studies is calculated as,  
\begin{equation}
    \text{Load} = \text{FTP Packet Size} \times \text{FTP Packet Arrival Rate} .
\end{equation}

To calculate the reliability figures in Fig.~\ref{fig:reliabilityscenario}, for each scenario, all the transmitted packets are recorded with their corresponding latency values. Then, amongst those packets, the ones with their latency smaller than or equal to $L$ ($100$ or $500$\:ms depending on the scenario) are counted. The final result is the ratio of the counted packets to the total number of packets. This calculation corresponds to~\eqref{eq:reliability}. 

Comparing the reliability figures of the scenarios in Table~\ref{tab:scenarios_and_reliability} with the ones in Fig.~\ref{fig:reliabilityscenario}, we observe that for \textit{Standard Data Communication}, the reliability is above the required $99.9$\%. Therefore, it is possible to meet the reliability requirements of this scenario by doing baseline handover parameter tuning. However, for the other scenarios, doing more than baseline handover tuning seems to be necessary, which we plan to pursue as part of our future work.  

\section{Future Outlook}
\label{sect:future_outlook}
Based on the results, we observed that we are able to meet the \textit{Standard Data Communications} scenario in Table~\ref{tab:scenarios_and_reliability}. Although optimizing for the handover parameters helped reduce the $99.9$th percentile of the latency distribution, it was not enough to push it under $100$\:ms to meet the requirements of the other scenarios. To lower the latency further, several other mobility procedures can be investigated. RACH-less LTM can be a candidate since the random access procedure occupies an important portion of the handover interruption time. DAPS handover can also be a candidate since it reduces the interruption time almost to zero. However, it requires complex hardware modifications at the UE side, which is not desirable by the UE vendors.

Another aspect that can be considered apart from improving mobility procedures is the cell deployment. In~\cite{etsi.103.554-2}, only $8$\:km ISD is considered for rural operation at $900$\:MHz. Therefore, in our simulation studies, the ISD is kept at $8$\:km. As a result of this, the gNodeBs are placed far from each other, creating difficulties for having a sufficient coverage at the cell edges. For lower power levels, this problem shows itself as radio link failures and for higher power levels, an extended cell edge together with a lot of ping-pong handovers, which we tried to reduce as much as possible. Of course, increasing the gNodeB transmission power excessively is not a solution considering the fact that maximum effective isotropic radiated power (EIRP) is limited to $63$\:dBm~\cite{etsi.103.554-2}. Cell densification can be performed either increasing the amount of cells at $900$\:MHz, or by introducing cells at $1900$\:MHz in between.

\section{Conclusion}
\label{sect:conclusion}
In this paper, we investigated the effect of baseline handover procedure on reliability in FRMCS and we tried to find the best parameter configuration that yields the lowest latency at $99.9$th percentile. We showed that with baseline handover procedure, we meet only one of the use cases, namely the \textit{Standard Data Communication} use case. We have further discussed the effect of A3 Event Threshold and TTT optimization to ping-pong handovers and how reducing them improves the $99.9$th percentile of the latency distribution. However, to meet the reliability requirements other scenarios, some different approaches are needed, as discussed in Section~\ref{sect:future_outlook}. 

Inline with these future directions discussed in Section~\ref{sect:future_outlook}, our focus will be on investigating the effect of different mobility procedures on reliability using the simulation setup described in Section~\ref{sect:simulation_scenario} as a reference. The effect of varying ISD and co-existence of cells operating with the center frequency of $1900$\:MHz alongside the ones operating on $900$\:MHz will also be investigated.
 
\bibliographystyle{IEEEtran}
\bibliography{bibliography}

\begin{thebibliography}{10}
\providecommand{\url}[1]{#1}
\csname url@samestyle\endcsname
\providecommand{\newblock}{\relax}
\providecommand{\bibinfo}[2]{#2}
\providecommand{\BIBentrySTDinterwordspacing}{\spaceskip=0pt\relax}
\providecommand{\BIBentryALTinterwordstretchfactor}{4}
\providecommand{\BIBentryALTinterwordspacing}{\spaceskip=\fontdimen2\font plus
\BIBentryALTinterwordstretchfactor\fontdimen3\font minus \fontdimen4\font\relax}
\providecommand{\BIBforeignlanguage}[2]{{%
\expandafter\ifx\csname l@#1\endcsname\relax
\typeout{** WARNING: IEEEtran.bst: No hyphenation pattern has been}%
\typeout{** loaded for the language `#1'. Using the pattern for}%
\typeout{** the default language instead.}%
\else
\language=\csname l@#1\endcsname
\fi
#2}}
\providecommand{\BIBdecl}{\relax}
\BIBdecl

\bibitem{eirene_frs}
{GSM-R Functional Group}, \emph{EIRENE Functional Requirements Specification V8.0.0}.\hskip 1em plus 0.5em minus 0.4em\relax International Union of Railways, 2023.

\bibitem{frmcs_5g_rail_uic_brochure}
``{FRMCS and 5G for rail: challenges, achievements and opportunities},'' \url{https://uic.org/IMG/pdf/brochure_frmcs_v2_web.pdf}, 2020, [Online; accessed 29-September-2023].

\bibitem{uic_frmcs}
``{Future Railway Mobile Communication System},'' \url{https://uic.org/rail-system/telecoms-signalling/article/frmcs/}, 2024, [Online; accessed 18-February-2024].

\bibitem{3gpp.22.289}
3GPP, ``{Mobile communication system for railways},'' {3GPP}, TS 22.289, 12 2019, v17.0.0.

\bibitem{5gsmart_deliverable_d1_5}
{5G-SMART}, \emph{D1.5 Evaluation of Radio Network Deployment Options}.\hskip 1em plus 0.5em minus 0.4em\relax 5G-SMART, 2023.

\bibitem{etsi.103.554-2}
ETSI, ``{Radio performance simulations and evaluations in rail environment; Part 2: New Radio (NR)},'' {ETSI}, TR 103 554-2, 02 2021, v1.1.1.

\bibitem{noh2020high}
G.~Noh, B.~Hui, and I.~Kim, ``High speed train communications in 5g: Design elements to mitigate the impact of very high mobility,'' \emph{IEEE Wireless Communications}, vol.~27, no.~6, pp. 98--106, 2020.

\bibitem{ali2020seamless}
W.~Ali, J.~Wang, H.~Zhu, and J.~Wang, ``Seamless mobility under a dedicated distributed antenna system for high-speed rail networks,'' \emph{IEEE Transactions on Vehicular Technology}, vol.~69, no.~12, pp. 15\,427--15\,441, 2020.

\bibitem{wu2022parameter}
C.~Wu, X.~Cai, J.~Sheng, Z.~Tang, B.~Ai, and Y.~Wang, ``Parameter adaptation and situation awareness of lte-r handover for high-speed railway communication,'' \emph{IEEE Transactions on Intelligent Transportation Systems}, vol.~23, no.~3, 2022.

\bibitem{chen2022handover}
Y.~Chen, K.~Niu, W.~Zhang \emph{et~al.}, ``Handover optimization algorithm based on t2rfs-fnn,'' \emph{Computational Intelligence and Neuroscience}, vol. 2022.

\bibitem{karamichailidis2021session}
P.~Karamichailidis \emph{et~al.}, ``Session management across heterogeneous wireless technologies in a rail transport environment,'' \emph{IEEE 5G FOR CONNECTED AND AUTOMATED MOBILITY (CAM)}, 2021.

\bibitem{cvetkovski2022railway}
D.~Cvetkovski \emph{et~al.}, ``Railway services support over a 5g infrastructure exploiting a multi-technology wireless transport network,'' in \emph{2022 IEEE FNWF}.\hskip 1em plus 0.5em minus 0.4em\relax IEEE, 2022, pp. 585--590.

\bibitem{5grail_deliverable_d6_2}
{The European Commission}, \emph{5G For Future Railway Mobile Communication System Deliverable D6.2}.\hskip 1em plus 0.5em minus 0.4em\relax The European Commission, 2023.

\bibitem{bovy2002analysis}
C.~Bovy, H.~Mertodimedjo, G.~Hooghiemstra, H.~Uijterwaal, and P.~Van~Mieghem, ``Analysis of end-to-end delay measurements in internet,'' in \emph{Proc. of the Passive and Active Measurement Workshop-PAM}, vol. 2002.\hskip 1em plus 0.5em minus 0.4em\relax sn, 2002.

\bibitem{atik2023reliabilityfrmcs}
D.~Atik, M.~Gursu, B.~Khodapanah, and W.~Kellerer, ``Reliability in future railway mobile communication systems,'' in \emph{2023 IEEE CSCN}.\hskip 1em plus 0.5em minus 0.4em\relax IEEE, 2023.

\bibitem{3gpp.38.331}
3GPP, ``{NR; Radio Resource Control (RRC); Protocol specification},'' {3GPP}, TS 38.331, 01 2024, v18.0.0.

\bibitem{3gpp.38.133}
3GPP, ``{NR; Requirements for support of radio resource management},'' {3GPP}, TS 38.133, 09 2023, v18.3.0.

\bibitem{3gpp.38.213}
3GPP, ``{NR; Physical layer procedures for control},'' {3GPP}, TS 38.213, 01 2024, v18.1.0.

\bibitem{freac}
F.~Abinader, C.~Rom, K.~Pedersen, S.~Hailu, and N.~Kolehmainen, ``System-level analysis of mmwave 5g systems with different multi-panel antenna device models,'' in \emph{2021 IEEE 93rd Vehicular Technology Conference (VTC2021-Spring)}.\hskip 1em plus 0.5em minus 0.4em\relax IEEE, 2021, pp. 1--6.

\bibitem{3gpp.38.901}
3GPP, ``{Study on channel model for frequencies from 0.5 to 100 GHz},'' {3GPP}, TR 38.901, 01 2024, v17.1.0.

\bibitem{ecc_20_02}
{ECC}, \emph{Harmonised use of the paired frequency bands 874.4-880.0 MHz and 919.4-925.0 MHz and of the unpaired frequency band 1900-1910 MHz for Railway Mobile Radio (RMR)}.\hskip 1em plus 0.5em minus 0.4em\relax European Conference of Postal and Telecommunications Administrations, 2020.

\bibitem{etsi.gsm.05.01}
ETSI, ``{Digital cellular telecommunications system (Phase 2+); Physical layer on the radio path; General description},'' {ETSI}, TS GSM 05.01, 03 1997, v5.2.0.

\bibitem{3gpp.36.889}
3GPP, ``{Study on Licensed-Assisted Access to Unlicensed Spectrum},'' {3GPP}, TR 36.889, 06 2015, v13.0.0.

\end{thebibliography}

\end{document}